\documentclass[10pt,aps,showpacs,a4paper,floatfix,twocolumn,tightenlines]{revtex4}
\usepackage{epsfig}
\usepackage{psfig}
\usepackage{graphicx}
\usepackage{graphics}
\usepackage{xspace}
\usepackage{amssymb}
\usepackage{latexsym}
\usepackage{natbib}
\usepackage{mathrsfs}
\newcommand{\inieq}{\begin{eqnarray}}            
\newcommand{\fineq}{\end{eqnarray}}            
\newcommand{\diff}{{\rm\,d}}                    
\newcommand{\half}{\textstyle{1\over 2}}

\def\p{\mbox{\boldmath $p$}}

\def\q{\mbox{\boldmath $q$}}

\def\ss{\mbox{\boldmath $\sigma$}}

\def\pf{\mbox{\boldmath $p^{ \prime}$}}
\def\pm{\mbox{\boldmath $p_{\mathrm m}$}}

\def\ep{\mbox{$e^{\prime}$}}

\begin{document}
\title{Relativistic approach to one nucleon knockout reactions} 

\author{Andrea Meucci} 
\altaffiliation[]{Talk presented by Andrea Meucci at the 
{\it IX Workshop on Theoretical Nuclear Physics in Italy}, Cortona, Italy, 9-12
October 2002 }
\author{Carlotta Giusti}
\author{Franco Davide Pacati }
\affiliation{Dipartimento di Fisica Nucleare e Teorica, 
Universit\`{a} di Pavia and \\
Istituto Nazionale di Fisica Nucleare, 
Sezione di Pavia, I-27100 Pavia, Italy}


\begin{abstract}
We develop a fully relativistic distorted wave impulse approximation 
model for electron- and photon-induced one proton knockout reactions. 
The relativistic mean field for the bound state
and the Pauli reduction for the scattering state are used, including a fully
relativistic electromagnetic current operator. Results for 
$^{16}$O$\left(e,e^{\prime}p\right)$ cross section and structure functions are
shown in various kinematic conditions and compared with nonrelativistic
calculations. Nuclear transparency calculations in a $Q^2$ range between $0.3$
and $1.8$ $($GeV$/c)^2$ are presented. 
Results for 
$^{16}$O$\left(\gamma,p\right)$ differential
cross sections are displayed in an energy range between 60 and 150 
MeV including two-body seagull contribution in the nuclear current. 
\end{abstract}


\maketitle


\section{Introduction}
\label{sec.int}

One nucleon knockout reactions are a primary tool to explore the 
single-particle
aspects of the nucleus. A long series of high-precision measurements at 
different energies and kinematics in a wide range of target nuclei
stimulated the
production of a considerable amount of theoretical calculations
\cite{book,kellyrep,RDWIA,RDWIA1,RDWIA2}. 
In the one-photon exchange approximation the coincidence cross section is given
by the contraction between the lepton tensor, completely determined by QED, and 
the hadron tensor, whose components depend on the transition matrix elements of
the nuclear current operator. 
In this note, the matrix elements of the nuclear 
current operator, i.e.,
\inieq
J^{\mu} = \langle \Psi_{\textrm {f}}\mid j^{\mu}\mid \Psi_{\textrm {i}}\rangle 
\ ,\label{eq.jmu}
\fineq
are calculated with relativistic wave functions for initial and scattering
states.
The bound state wave functions, i.e., 
\inieq
\Psi_{\textrm {i}} = \left(\begin{array}{c} u_{\textrm {i}}  \\ 
 v_{\textrm {i}} \end{array}\right) \ , \label{eq.bwf}
\fineq
are given by the Dirac-Hartree solution of a relativistic Lagrangian
containing scalar and vector potentials \cite{adfx}. 
The ejectile wave function is written in terms of its positive energy
component following the direct Pauli reduction scheme, i.e.,
\inieq
\Psi_{\textrm {f}} = \left(\begin{array}{c} \Psi_{\textrm {f}+} \\ 
\frac {\ss\cdot\p'}{M+E'+S-V}
        \Psi_{\textrm {f}+} \end{array}\right) \ ,
\fineq
where $S=S(r)$ and $V=V(r)$ are the scalar and vector potentials for the nucleon
with energy and momentum $E'$ and $\pf$~\cite{cooper}. 
The upper component, $\Psi_{\textrm {f}+}$, is related to a
Schr\"odinger equivalent wave function, $\Phi_{\textrm{f}}$, by the Darwin 
factor, $D(r)$, i.e.,
\inieq
\Psi_{\textrm {f}+} &=& \sqrt{D(r)}\Phi_{\textrm{f}} \ , \\
D(r) &=& 1 + \frac{S-V}{M+E'} \ . \label{eq.darw}
\fineq
$\Phi_{\textrm{f}}$ is a two-component wave function which is solution of a 
Schr\"odinger
equation containing equivalent central and spin-orbit potentials obtained from
the scalar and vector potentials.
The most common current conserving prescriptions, i.e., cc1, cc2, and 
cc3 \cite{defo}, are used for the one-body current $j^{\mu}$.

\section{The $\left(e,e^{\prime}p\right)$ reaction} \label{sec.eep}
The coincidence cross section of the $\left(e,\ep p\right)$ reaction can be 
written in terms of four response functions,
$R_{\lambda\lambda^{\prime}}$, as
\inieq
\sigma  = 
K\ \left\{v _{L}
R_{L} +  v_{T}R_{T}+v_{LT}R_{LT}\cos\left(\vartheta\right) \right. \nonumber \\
\left. +v_{TT}R_{TT}\cos\left(2\vartheta\right)\right\} \ ,  
\label{eq.fcs}
\fineq
where $K$ is a kinematic factor, and $\vartheta $ is the out-of-plane 
angle between the electron scattering plane and the $(\q, \pf)$ plane. The 
coefficients $v_{\lambda\lambda'}$ are obtained from the lepton tensor
components and depend only upon the electron kinematics \cite{book,kellyrep}.

The response functions are given by bilinear combinations of the nuclear current
components, i.e.,
\inieq
R_{L} &\propto &\langle J^0 \left(J^0\right)^{\dagger }\rangle \ , \nonumber \\
R_{T} &\propto &\langle J^x \left(J^x\right)^{\dagger} \rangle +
          \langle J^y \left(J^y\right)^{\dagger} \rangle \ , \nonumber \\
R_{LT} &\propto &-2\ {\rm Re}
   \left[\langle J^x \left(J^0\right)^{\dagger }\rangle\right] \ , \nonumber \\
R_{TT} &\propto & \langle J^x \left(J^x\right)^{\dagger} \rangle -
          \langle J^y \left(J^y\right)^{\dagger} \rangle \ ,
\fineq
where $\langle\cdots\rangle$ means that average over the initial and sum over
the final states is performed fulfilling energy conservation. 


\begin{figure}[ht]
\includegraphics[scale=.55]{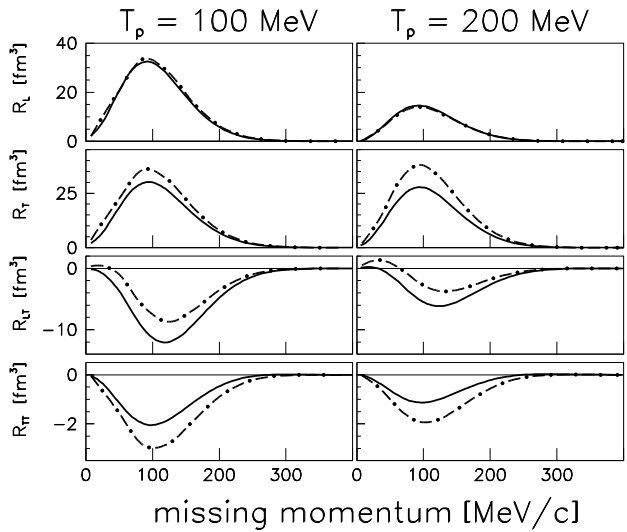}
\caption {The response functions for the $^{16}$O$(e,e'p)^{15}$N
reaction at $T_p = 100$ (left panel) and $200$ MeV (right panel)  
in the center-of-mass system in constant $(\q,\omega)$ kinematics. 
Solid lines are the relativistic results, dot-dashed lines are the 
nonrelativistic results. }
\label{fig1}
\end{figure}

The comparison between our relativistic distorted wave impulse approximation 
(RDWIA) results and the nonrelativistic results
\cite{book} 
is shown in Fig. \ref{fig1}\ for the response functions of the 
$^{16}$O$(e,e'p)^{15}$N$_{\mathrm{g.s}}$ reaction in constant $\left(\q,
\omega\right)$ kinematics at two different values of the 
proton energy, i.e., $T_p=$ 100 and $ 200$ MeV \cite{mgp}. The cc2 prescription 
has been 
used. The differences rapidly increase with the energy, and the
relativistic results are smaller than the nonrelativistic ones. This outcome is
well-known and it is essentially due to the $\sqrt{D}$ factor of Eq.
\ref{eq.darw} and to the relativistic normalization factor $(M+E')/(2E')$. Small
differences are obtained for the longitudinal response function $R_L$. On the
contrary, a visible reduction is found for the transverse response $R_T$, even
at $T_p = 100$ MeV. Large differences are generally found for the 
longitudinal-transverse interference response $R_{LT}$. The combined
relativistic effects on $R_T$ and $R_{LT}$ are responsible for the different
asymmetry in the cross section. The transverse-transverse interference 
response $R_{TT}$ is much smaller than the other response functions and gives
only a negligible contribution to the cross section.


\begin{figure}[ht]
\includegraphics[scale=.45]{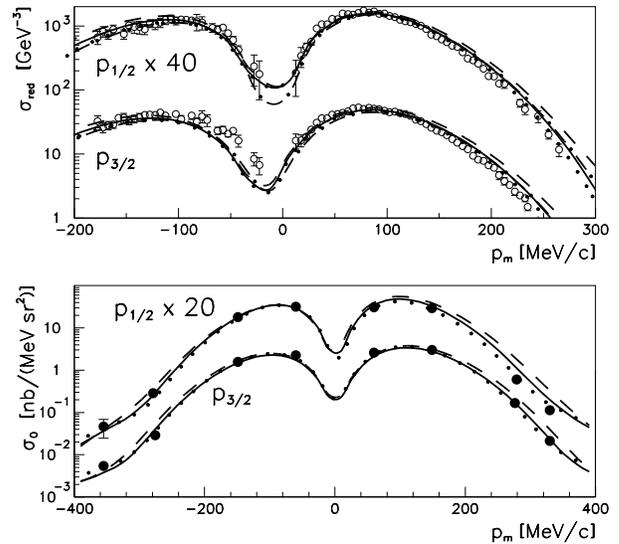}
\caption {The cross section for the $^{16}$O$(e,e'p)^{15}$N
reaction at $T_p = 90$ MeV in parallel kinematics (upper panel) \cite{nikhef},
and at $Q^2=0.8$ (GeV/$c)^2$ in 
constant $(\q,\omega)$ kinematics \cite{e89003} (lower panel). Data 
for the $p\half$ state have been multiplied by 40 and 20, respectively.
Dashed, solid, and dotted lines 
represent the result of the RDWIA approach with cc1, cc2, cc3 off-shell 
prescriptions, respectively. Dot-dashed lines in the upper panel are the 
nonrelativistic results. }
\label{fig2}
\end{figure}

In the upper panel of
Fig.~\ref{fig2}, the reduced cross section section data measured at NIKHEF
in parallel kinematics at a constant $T_p \simeq$ 90 MeV in the 
center-of-mass system \cite{nikhef} are compared with
our calculations \cite{mgp,rmd}.  
The relativistic curves have been rescaled by the spectroscopic factors 
$Z_{p1/2}= 0.71$ and $Z_{p3/2}=0.60$, while the nonrelativistic ones by 
$Z_{p1/2}= 0.64$ and $Z_{p3/2}=0.54$.
It is apparent that, at low energy, relativistic and nonrelativistic
calculations are almost equivalent in comparison with data. 
However, the fact that the relativistic 
calculations are normalized to
experimental data with a closer to 1 spectroscopic factor, 
seems to indicate that a relativistic treatment should be preferred with 
respect to a nonrelativistic one. 
The sensitivity to the off-shell
ambiguity in the electromagnetic current operator is relatively weak and 
remains within a range of about 10\%.

In the lower panel, the same reaction is considered
at the Jlab constant $(\q, \omega)$ kinematics with $Q^2 = 0.8$ 
(GeV/$c$)$^2$~\cite{e89003}. 
The spectroscopic factors are the same as in the upper
panel. The agreement with data is still very good. We remark that, at this
energy, a relativistic treatment is necessary to correctly describe the data.

\section{Nuclear transparency} \label{sec.transpa}

\begin{figure}[ht]
\includegraphics[scale=.48]{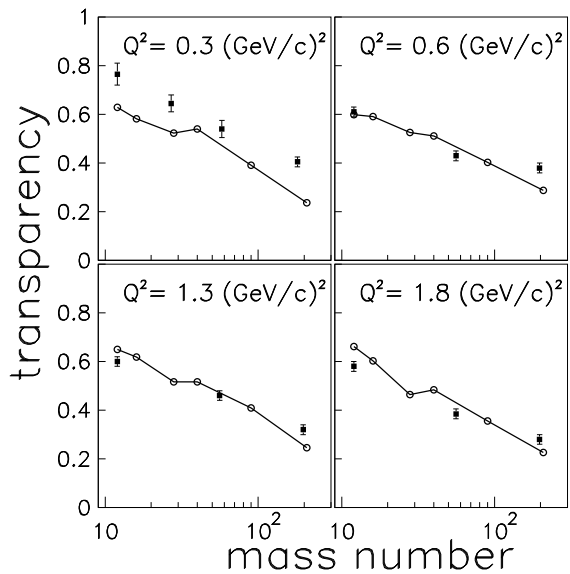}
\caption {The nuclear transparency for the quasifree 
A$\left(e,\ep p\right)$ reaction as a function of the mass number for $Q^2$
ranging from $0.3$ to $1.8$ (GeV$/c)^2$. Calculations have been performed for
selected closed shell or subshell nuclei with mass numbers indicated by open 
circles. The data at $Q^2 = 0.3$ (GeV$/c)^2$ are from Ref.~\cite{garino}. The 
data at $Q^2 = 0.6, 1.3$, and $1.8$ (GeV/$c)^2$ are from Ref.~\cite{abbott}.
\label{fig3}}

\end{figure}

The nuclear transparency can be intuitively defined as the
ratio of the measured to the plane wave cross section. The transparency can be
used to refine our knowledge of nuclear medium effects and to look for deviation
from conventional nuclear physics, such as the Color Transparency
(CT) effect \cite{garino,abbott,oneill}. If the CT effect switches on as 
$Q^2$ increases, then the nuclear transparency should be enhanced towards unity.
 
Several measurements of the nuclear transparency in $\left(e,\ep p\right)$ 
have been carried out in the past, but they did not show any evidence for the 
onset of CT in a $Q^2$ range up to 8.1 (GeV/$c$)$^2$.

We define nuclear transparency as
\begin{eqnarray}
T = \frac {\int_V \diff E_{\mathrm m} \diff\pm~\sigma_{DW} \left( 
E_{\mathrm m}, \pm, \pf\right)}{\int_V \diff E_{\mathrm m}\diff\pm~ 
\sigma_{PW} \left(E_{\mathrm m}, \pm\right)} \ , \label{eq.transpa}
\end{eqnarray} 
where $\sigma_{DW}$ is the distorted wave cross section and $\sigma_{PW}$ is 
the plane wave one. The integration is performed upon the space phase 
volume $V$.

In Fig.~\ref{fig3} our RDWIA results for nuclear transparency, calculated with
the cc2 prescription for the nuclear current, are shown \cite{meu}. 
The $Q^2$ of the 
exchanged photon is taken between $0.3$ (GeV/$c)^2$ and $1.8$ (GeV/$c)^2$ in 
constant $\left(\q, \omega\right)$ kinematics. Calculations have been performed 
for selected closed shell or subshell nuclei. The agreement with the data is 
rather satisfactory. At 
$Q^2 = 0.3$ (GeV/$c)^2$ our results lie below the data, while at 
$Q^2 = 0.6, 1.3$, and $1.8$ (GeV/$c)^2$  they are closer to the data and fall 
down only for higher mass numbers.
The discontinuities of the shell structure clearly appear in the changes in
shape of the A-dependent curves.

\section{Photoreactions} \label{sec.photo}

In case of an incident photon with energy $E_{\gamma}$, the 
$\left(\gamma,N\right)$ cross section can be written in terms of the pure
transverse response, i.e.,
\inieq
\sigma_{\gamma} = K'\  R_{TT} \ ,  
\label{eq.gcs}
\fineq
where $K'$ is a kinematic factor.

\begin{figure}[ht]
\centerline{\includegraphics[scale=.48]{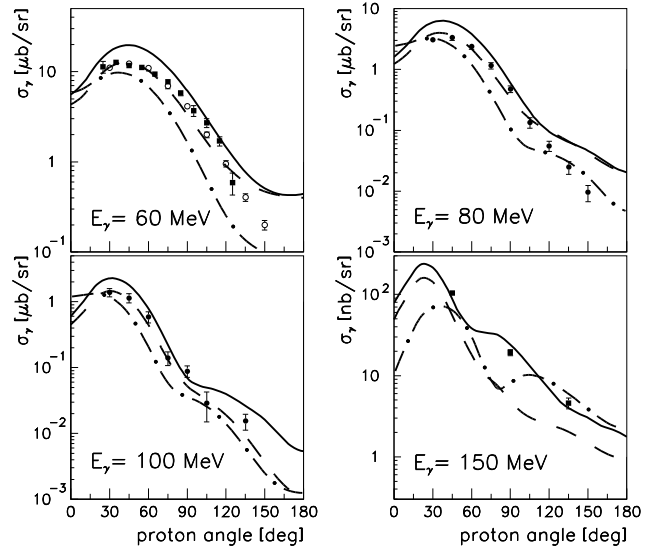}}
\caption {The cross section for the $^{16}$O$(\gamma ,p)^{15}$N reaction as a 
function of the proton scattering angle at a photon energy
ranging from $60$ to $150$ MeV. 
The data at $60$ MeV are from Refs. \cite{miller,findlay}. The 
data at $80$ and $100$ MeV are from Ref. \cite{findlay}. The 
data at $150$ MeV are from Ref. \cite{leitch}. Solid lines represent the
relativistic calculations with the inclusion of the seagull current, dashed
lines the relativistic results with one-body current only, and dot-dashed lines
are the one-body nonrelativistic results.
\label{fig4}}

\end{figure}

The comparison between relativistic and nonrelativistic results is shown in
Fig.~\ref{fig4} for the cross section of the
$^{16}$O$\left(\gamma,p\right)^{15}$N$_{\textrm{g.s.}}$ reaction at
photon energy ranging from $60$ to $150$ MeV \cite{mgpho}. The cc2 current has 
been used and the same spectroscopic
factor $Z(p\half)=0.71$ of $\left(e,e^{\prime}p\right)$ data has been applied.

We see that the differences between the nonrelativistic calculations and the
relativistic ones with the cc2 prescription are sensible at all energies. 
The nonrelativistic results are always smaller than the
data. On the contrary, the 
relativistic results with the one-body current are generally closer to the 
data and well reproduce the 
magnitude and shape, at least at low energies.    
For higher energies, the relativistic results fall below the data and the 
discrepancies increase with the proton angle.

As a first step to study the role of meson exchange currents in photoreactions,
the results with the inclusion of the seagull contribution in the current are
also shown in Fig. \ref{fig4} \cite{mgpseag}.
The pure contribution of the two-body term 
is one order of magnitude lower than the one-body one, but their interference 
is large. Thus, the total result is enhanced above the
data and the shape is slightly affected. The seagull contribution 
is sizable but less than in previous nonrelativistic
calculations \cite{book}. On the other hand, in nonrelativistic calculations,
the pion-in-flight diagram reduces the effects of the seagull current, while 
the $\Delta$ 
current is important only with increasing photon energies. 
If these results 
were confirmed in relativistic calculations, the pion-in-flight term would
reduce the contribution of seagull and bring the calculated cross 
section in Fig.~\ref{fig4} closer to the one-body results and also to the data.

\end{document}